\documentclass[prl, aps, twocolumn, floatfix, showpacs]{revtex4-1}
\usepackage{graphicx, amsmath, amssymb, times}

\begin{document}
\title{Stripe-ordered superfluid and supersolid phases in attractive Hofstadter-Hubbard model}
\author{M. Iskin}
\affiliation{
Department of Physics, Ko\c c University, Rumelifeneri Yolu, 34450 Sar{\i}yer, Istanbul, Turkey.
}
\date{\today}

\begin{abstract}

We use microscopic Bogoliubov-de Gennes formalism to explore the 
ground-state phase diagram of the single-band attractive Hofstadter-Hubbard 
model on a square lattice. We show that the interplay between the Hofstadter 
butterfly and superfluidity breaks spatial symmetry, and gives rise to 
stripe-ordered superfluid and supersolid phases in large parameter spaces. 
We also discuss the effects of a trapping potential and comment on the viability 
of observing stripe-ordered phases with cold Fermi gases.

\end{abstract}
\pacs{03.75.Ss, 03.75.Hh, 67.85.-Lm, 67.85.-d, 67.80.kb}
\maketitle

\textit{Introduction.}
The so-called Hofstadter butterfly (HB) refers to the fractal energy spectrum of a 
quantum particle that is confined to move on a two-dimensional tight-binding periodic
lattice under the influence of a uniform magnetic flux~\cite{hofstadter76}. 
There are only two length scales in this simple problem, \textit{i.e.}, the 
lattice spacing $\ell$ and cyclotron radius $\ell_B$, and their competition 
produce one of the first quantum fractals discovered in physics. 
This self-similar spectrum exhibits a complex pattern of sub-bands 
and mini-gaps as a function of $\ell/\ell_B$, but despite all efforts since 
its prediction, the limited experimental control over this ratio 
$(\ell/\ell_B \ll 1$ even for the largest attainable magnetic fields)
has hampered the development of techniques that could probe its effects 
in natural solid-state crystals. It was only last year that it became 
possible to observe signatures of this spectrum in graphene-based 
materials with artificially-engineered superlattices under real magnetic 
fields, however, its full structure largely remains uncharted 
territory~\cite{dean13, ponomarenko13, kuhl}.

In addition, inspired by the recent realisation of artificial gauge 
fields~\cite{galitski13, dalibard11, chen12, wang12,cheuk12,qu13,fu13, williams13}, 
the quest for the Hofstadter spectrum and related phenomena have been 
revitalised in the cold-atom 
community~\cite{garcia12, struck12, kennedy13, aidelsburger13, miyake13, chin13, cocks12, wang14}. 
For instance, by engineering spatially-dependent complex tunneling 
amplitudes with laser-assisted tunneling and a potential energy gradient, 
two independent research groups have recently reported compelling 
evidence for the realisation of the Hofstadter-Harper Hamiltonian with 
neutral rubidium atoms that are loaded into laser-induced periodic
potentials~\cite{aidelsburger13, miyake13, chin13}. Thanks to their greater 
promise of engineering fully-tuneable many-body Hamiltonians on-demand, 
even though atomic systems are considered as one of the top candidates for 
much-desired quantum simulators, currently attainable temperatures in these 
works are not low enough for resolving the fractal structure of 
the HB. Nevertheless, captivated by these flourishing 
efforts, here we study the interplay between the HB, 
strong interactions and Zeeman field, and explore ground-state phases 
of the attractive Hofstadter-Hubbard model~\cite{zhai10}. 

Our main results are highlighted as follows. We find that the cooperation between 
the HB and superfluid (SF) order breaks the spatial symmetry 
of the system. This is in accordance with a recent work showing that superfluidity 
necessarily breaks translation symmetry in repulsive Hofstadter-Bose-Hubbard 
model~\cite{powell10}. In addition, we show that the phase diagrams are 
dominated by stripe-ordered SF and supersolid (SS) 
phases which are characterised by their coexisting pair-density (PDW), 
charge-density (CDW) and/or spin-density wave (SDW) 
orders~\cite{zhaiweinote, zhai10, wei12, agterberg08}. 
While these phases share some characteristic features of the long-sought 
Fulde-Ferrel-Larkin-Ovchinnikov (FFLO or LOFF) phase~\cite{FF64,LO65}, 
they are not driven by the Zeeman field and have an entirely new physical
mechanism. Given that FFLO-like phases are of high-demand in condensed-matter, 
nuclear and elementary-particle physics~\cite{casalbuoni04, prokofev12, anglani14, kivelson03, wu11}, 
our findings allude a new route towards creating them by loading neutral 
atomic Fermi gases on laser-induced optical lattices under laser-induced 
gauge fields.

\textit{Bogoliubov-de Gennes (BdG) Formalism.}
To achieve these results, we solve the single-band attractive Hofstadter-Hubbard 
Hamiltonian on a square lattice within the mean-field approximation for the 
on-site interaction term, \textit{i.e.},
%
%%
%\begin{align}
%\label{eqn:hamiltonian}
%H = &- \sum_{ij\sigma} t_{ij} a_{i\sigma}^\dagger a_{j\sigma}
%-\sum_{i\sigma} \mu_{i \sigma} a_{i\sigma}^\dagger a_{i\sigma} \nonumber \\
%&+ \sum_{i} \left( \Delta_i a_{i\uparrow}^\dagger a_{i\downarrow}^\dagger 
%+ \Delta_i^* a_{i\downarrow} a_{i\uparrow} + \frac{|\Delta_i|^2}{g} \right),
%\end{align}
%%
%
$
H = - \sum_{ij\sigma} t_{ij} a_{i\sigma}^\dagger a_{j\sigma}
-\sum_{i\sigma} \mu_{i \sigma} a_{i\sigma}^\dagger a_{i\sigma}
+ \sum_{i} \left( \Delta_i a_{i\uparrow}^\dagger a_{i\downarrow}^\dagger 
+ \Delta_i^* a_{i\downarrow} a_{i\uparrow} + \frac{|\Delta_i|^2}{g} \right),
$
where $a_{i\sigma}^\dagger$ ($a_{i\sigma}$) creates (annihilates) a $\sigma$ fermion
on site $i$, $t_{ij}$ is the tunneling (hopping) matrix element, and
$\mu_{i \uparrow} = \mu - gn_{i\downarrow} - V_i + h$ 
and
$\mu_{i \downarrow} = \mu - gn_{i\uparrow} - V_i - h$ 
are effectively the local chemical potentials in the presence of Hartree shifts,
confining potential $V_i$ and an out-of-plane Zeeman field $h \ge 0$. 
The complex hopping matrix $t_{ij}$ is assumed to connect only the 
nearest-neighbor sites, \textit{i.e.}, $t_{ij} = t e^{i\theta_{ij}}$ with $t \ge 0$ for $i$ and $j$ 
nearest neighbors and $0$ otherwise, and the phase
$
\theta_{ij} = (1/\phi_0) \int_{\mathbf{r}_i}^{\mathbf{r}_j} \mathbf{A}(\mathbf{r}) \cdot d\mathbf{r}
$
takes the effects of real external magnetic fields (or artificial gauge fields) into 
account via the Peierls substitution. Here, $\mathbf{A}(\mathbf{r})$ is the 
corresponding vector potential and $\phi_0 = 2\pi \hbar/e$ is the magnetic flux 
quantum. The remaining terms involve the complex SF order parameter
$
\Delta_i = g \langle a_{i\uparrow} a_{i\downarrow} \rangle,
$
where $g \ge 0$ is the strength of the on-site density-density interaction and 
$\langle \cdots \rangle$ is a thermal average.

This microscopic Hamiltonian can be diagonalised via the Bogoliubov-Valatin 
transformation, 
\textit{i.e.},
$
a_{i\sigma} = \sum_m (u_{mi\sigma} \gamma_{m\sigma} 
- s_\sigma v_{mi\sigma}^* \gamma_{m,-\sigma}^\dagger),
$
where $\gamma_{m\sigma}^\dagger$ ($\gamma_{m\sigma}$) creates (annihilates) 
a pseudo-spin $\sigma$ quasiparticle with energy $\epsilon_{m}^\sigma$ and wave 
functions $u_{mi\sigma}$ and $v_{mi\sigma}$, and the resultant BdG equations can be 
compactly written as,
\begin{equation}
\label{eqn:bdg}
\sum_j 
\left( \begin{array}{cc}
-t_{ij} - \mu_{i\uparrow} \delta_{ij} & \Delta_i \delta_{ij} \\
\Delta_i^*\delta_{ij} & t_{ij}^* + \mu_{i\downarrow} \delta_{ij}
\end{array} \right)
\varphi_{mj}^\sigma
 = s_\sigma \epsilon_{m}^\sigma \varphi_{mi}^\sigma.
\end{equation}
Here, $\delta_{ij}$ is the Kronecker delta,
$
\varphi_{mi}^{\uparrow} = (u_{mi\uparrow}^*, v_{mi\downarrow}^*)^\dagger
$
and
$
\varphi_{mi}^{\downarrow} = (v_{mi\uparrow}, -u_{mi\downarrow})^\dagger
$
are the corresponding eigenfunctions for $\epsilon_{m}^\sigma \ge 0$ eigenvalue,
and $s_\uparrow = +1$ and $s_\downarrow = -1$.
Note that the BdG equations are invariant under the transformation
$v_{mi\uparrow} \to u_{mi\uparrow}^*$, $u_{mi\downarrow} \to -v_{mi\downarrow}^*$ 
and $\epsilon_{m\downarrow} \to -\epsilon_{m\uparrow}$, and therefore, it is 
sufficient to solve only for $u_{mi} \equiv u_{mi\uparrow}$, $v_{mi} \equiv v_{mi\downarrow}$ 
and $\epsilon_m \equiv \epsilon_{m}^\uparrow$ as long as all solutions with 
positive and negative $\epsilon_m$ are kept.
For instance, $\Delta_i$ is given by
$
\Delta_i = - g\sum_m u_{mi} v_{mi}^* f(\epsilon_m),
$
where $f(x) = 1/[e^{x/(k_B T)} + 1]$ is the Fermi function with $k_B$ the 
Boltzmann constant and $T$ the temperature, and it has to be determined
self-consistently with $\mu$ and $h$ such that the total number of 
$\sigma$ fermions satisfies $N_\sigma = \sum_i n_{i\sigma}$. 
Here,
$
0 \le n_{i \sigma} = \langle a_{i\sigma}^\dagger a_{i\sigma} \rangle \le 1
$
is the number of $\sigma$ fermions on site $i$ given by
$
n_{i\uparrow} = \sum_m |u_{mi}|^2 f(\epsilon_m)
$
and
$
n_{i\downarrow} = \sum_m |v_{mi}|^2 f(-\epsilon_m).
$
When $\theta_{ij} = 0$, it is generally accepted that the mean-field description 
provides a qualitative understanding of the ground state~\cite{fhreview, stoof}, 
and here we include both $\theta_{ij}$ and $V_i$ exactly into the mean-field 
formalism without relying on further approximations. 

Without loosing generality, we choose Landau gauge for the vector potential, 
\textit{i.e.}, $\mathbf{A}(\mathbf{r}) \equiv (0,B x,0)$, leading to a uniform magnetic flux 
$\Phi = B \ell^2$ (per unit cell), where $\ell$ is the lattice spacing.
Denoting ($x,y$) coordinates of site $i$ by ($n\ell, m\ell$), 
this gauge simply implies $\theta_{ij} = 0$ and $\theta_{ij} = \pm 2\pi n\phi$ 
for links along the $x$ and $y$ directions, respectively,
where $\phi = \Phi/(2\pi \phi_0)$ characterises the competition between the 
lattice spacing $\ell$ and magnetic length scale $\ell_B = \sqrt{\hbar/(eB)}$. 
In this paper, we only consider $\phi = p/q$ ratios, where $p$ and $q$ are 
co-primes, for which the exact non-interacting single-particle excitation 
spectrum $\varepsilon(\phi)$ vs. $\phi$ is known as the HB~\cite{hofstadter76}
While $\phi$ remains $\ll 1$ for typical electronic crystals, even for the 
largest $B$ field that is attainable in a laboratory, it can in principle be 
tuned at will for atomic systems via artificial gauge fields. 
The fractal structure of the spectrum is expected to have drastic effects 
on the many-body problem which is our main motivation.

\textit{Ground-State Phases.}
In order to explore the possible phases, let us set $V_i = 0$ and consider a 
uniform $45\ell \times 45\ell$ square lattice, which is large enough to construct the 
thermodynamic phase diagrams for $\phi = \lbrace 0, 1/6, 1/4 \rbrace$. 
The Hartree shifts are neglected for simplicity~\cite{Hartree}. It turns out that the BdG 
equations allow for multiple solutions, especially for the polarised many-body 
phases, and therefore, it is essential to verify the (meta)stability of the 
solutions~\cite{ms}.

\begin{table}
\begin{tabular}[c]{ccccc}
\hline
Phase & $|\Delta_i|$ & $n_{i\uparrow}+n_{i\downarrow}$ & $n_{i\uparrow}-n_{i\downarrow}$ \\
\hline
U-SF & Uniform & Uniform & 0 \\
S-SF & PDW & 1 & 0 \\
S-SS & PDW & CDW & 0\\
\hline
\end{tabular}
\caption{\label{tab:sf}
The uniform-SF (U-SF), striped-SF (S-SF) and striped-SS (S-SS) phases can
be distinguished by their coexisting order parameters.}
\end{table}

\begin{figure}[htb]
\centerline{\scalebox{0.55}{\includegraphics{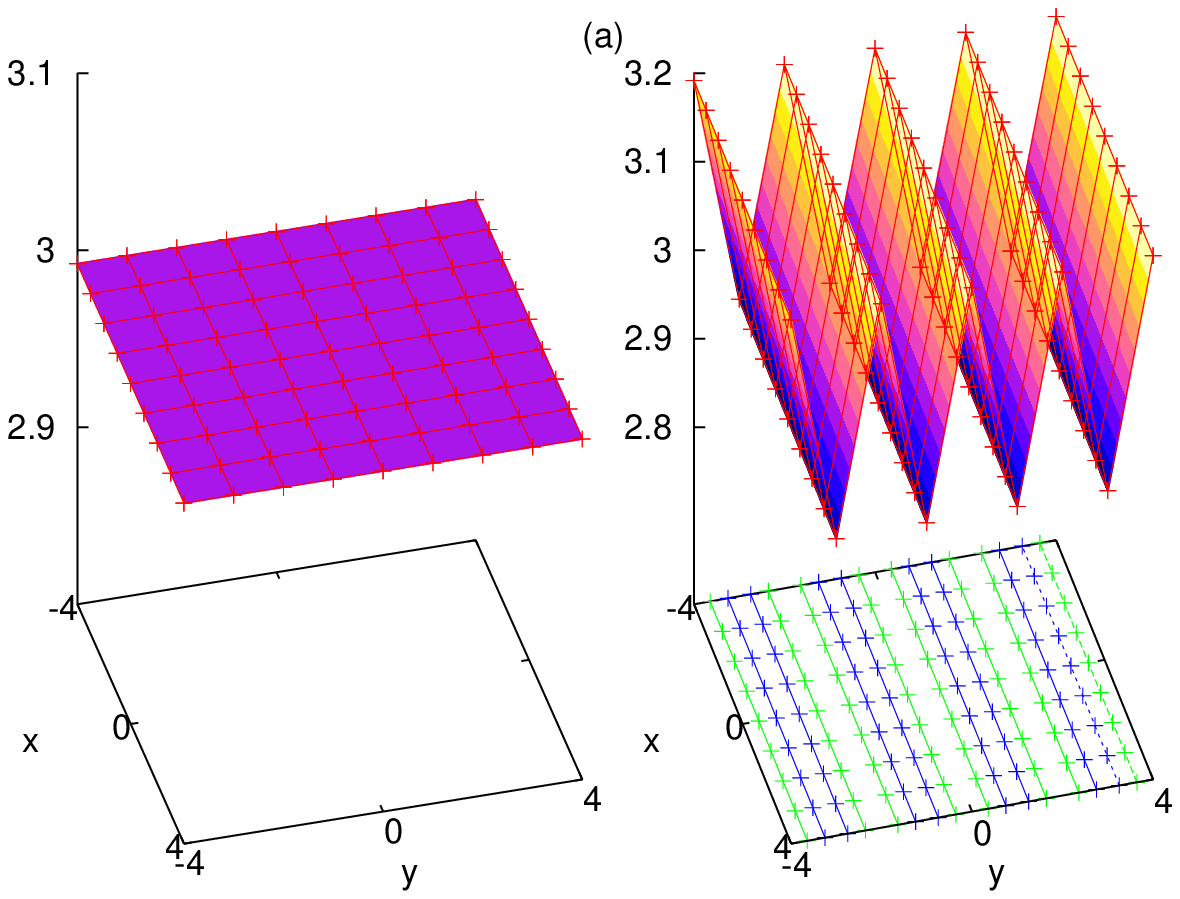}}}
\centerline{\scalebox{0.55}{\includegraphics{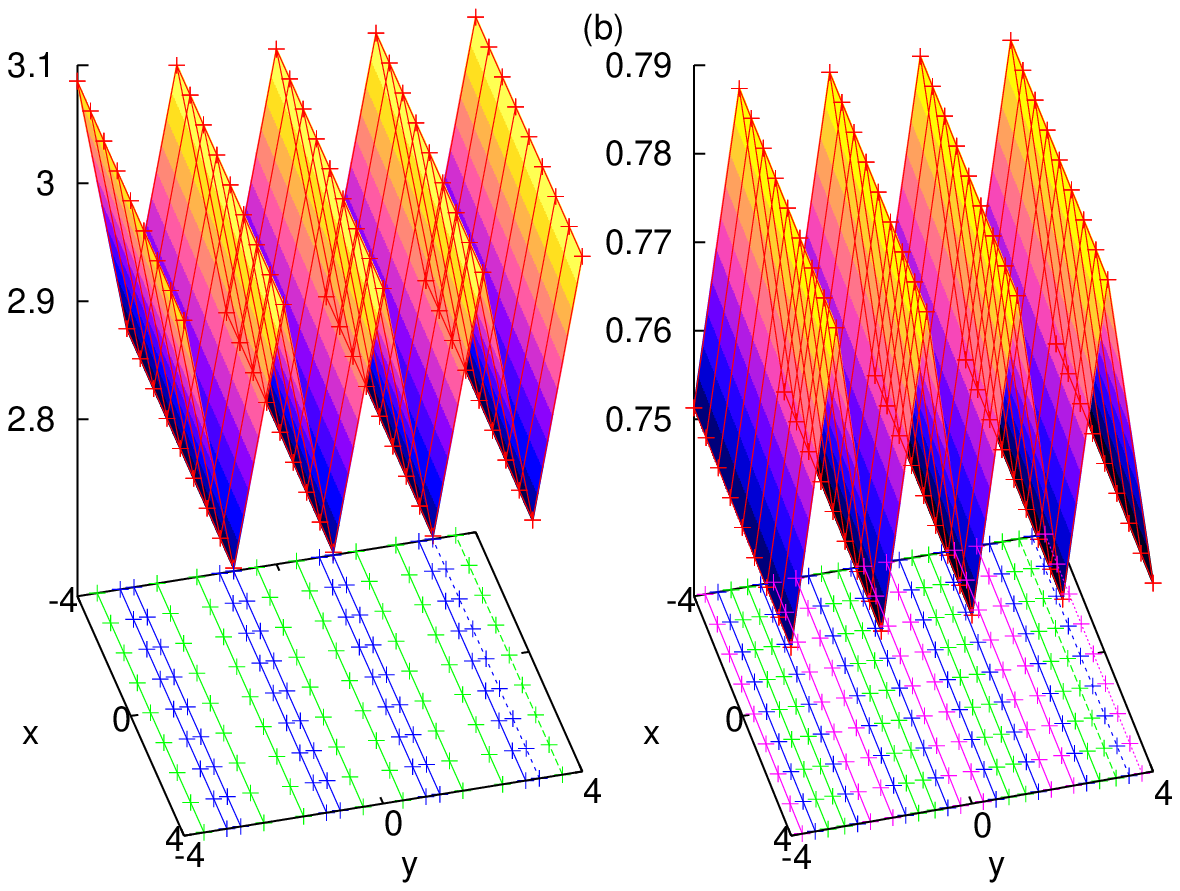}}}
\caption{\label{fig:S} (Color online)
(a) Typical $|\Delta_i|/t$ profiles are shown for the U-SF (left) and 
S-SF phases (right), where $\phi_\uparrow = \phi_\downarrow = 0$
and $\phi_\uparrow = \phi_\downarrow = 1/4$, respectively, and 
$\mu = 0$ corresponding to a uniformly half-filled lattice.
(b) Typical $|\Delta_i|/t$ (left) and $n_{i\uparrow} + n_{i\downarrow}$ (right) 
profiles are shown for the S-SS phase, 
where $\phi_\uparrow = \phi_\downarrow = 1/4$ and $\mu = -t$.
Here, $(x,y)$ are in units of $\ell$, and we set $h = 0$ and $g = 7t$ in 
all figures.
}
\end{figure}

Depending on the spatial profiles of $|\Delta_i|$, $n_{i\uparrow}$ and 
$n_{i\downarrow}$, we distinguish the single-particle phases from the 
many-body ones using the following scheme. 
When $h/g$ is sufficiently high that $\Delta_i \to 0$ 
($|\Delta_i| < 10^{-3}t$ in numerics), the ground state can be 
a $\sigma$-VAC phase which is a vacuum of $\sigma$ component,
a $\sigma$-I$(m/n)$ phase which is a band insulator of $\sigma$ 
component with uniform $n_{i\sigma} = m/n$,
a $\sigma$-N phase which is a normal $\sigma$ component,
or an $\uparrow \downarrow$-PN phase which is a polarised normal 
mixture of $\uparrow$ and $\downarrow$ components. 
On the other hand, when $h/g$ is sufficiently low, the ground state can be 
characterised according to Table~\ref{tab:sf}, and typical $|\Delta_i|$ 
and $n_{i\uparrow} + n_{i\downarrow}$ profiles are illustrated in 
Fig.~\ref{fig:S} for the uniform-SF, striped-SF and striped-SS phases.

\begin{figure}[htb]
\centerline{\scalebox{0.3}{\includegraphics{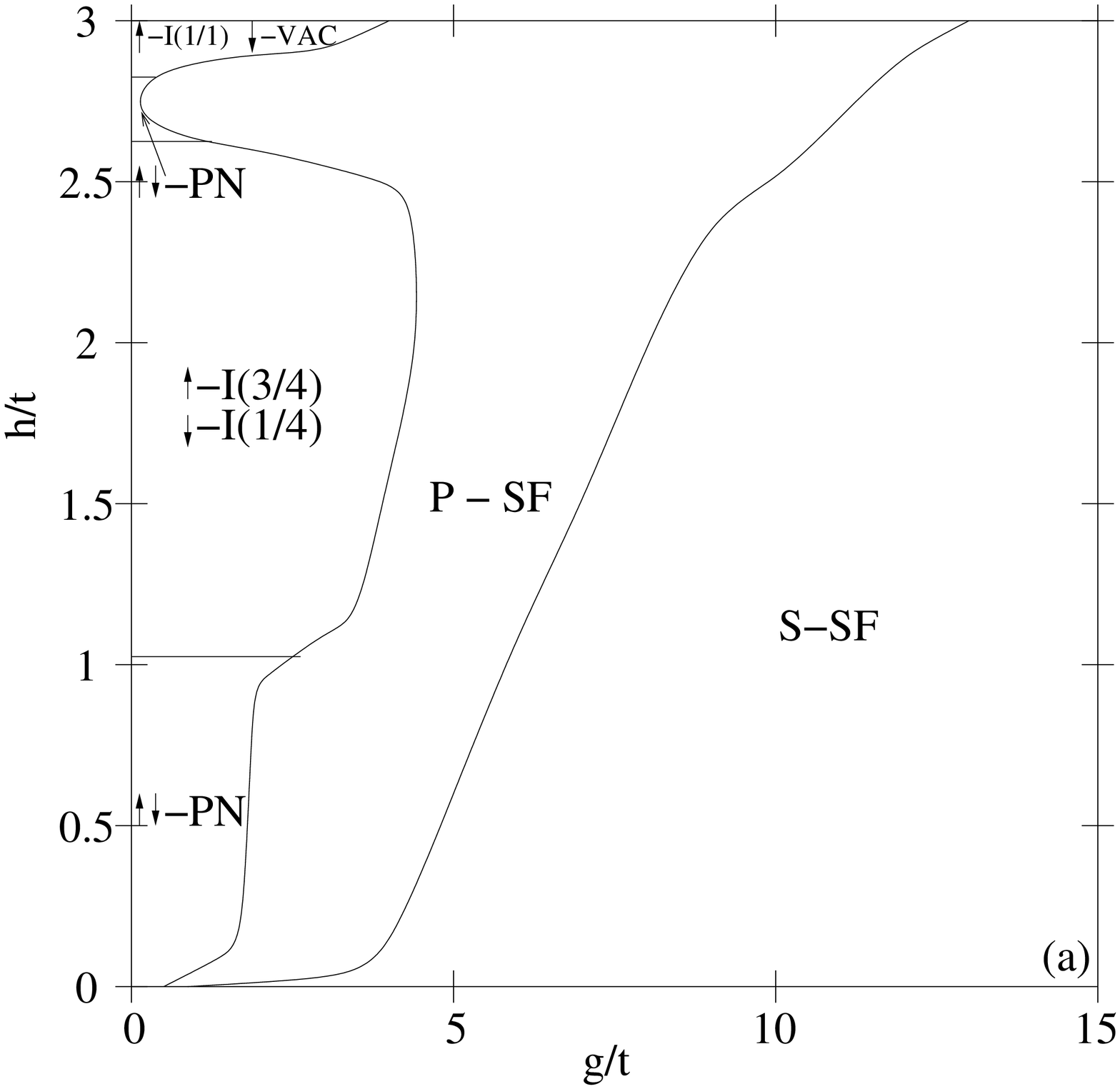}}}
\centerline{\scalebox{0.3}{\includegraphics{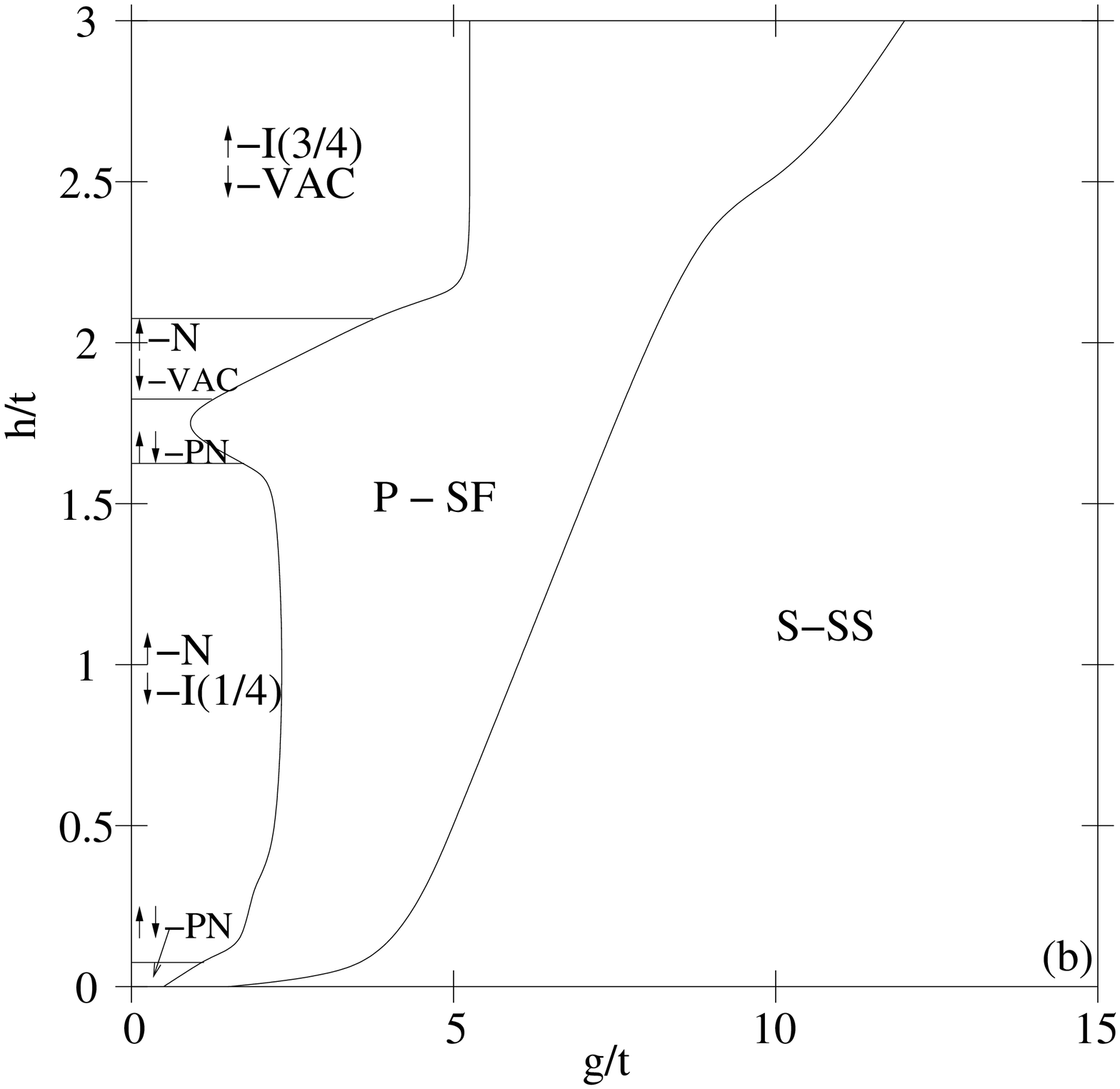}}}
\caption{\label{fig:pq14} (Color online) 
The ground-state phase diagrams are shown for $\mu = 0$ in (a) and $\mu = -t$ in (b),
when $\phi = p/q = 1/4$.}
\end{figure}

\textit{Thermodynamic Phase Diagrams.}
In Fig.~\ref{fig:pq14}, we present the $\phi = 1/4$ phase diagrams 
for $\mu = 0$ in Fig.~\ref{fig:pq14}(a) and $\mu = -t$ in Fig.~\ref{fig:pq14}(b).
The $\mu = 0$ case is very special since it corresponds to a half-filled 
lattice with particle-hole symmetry, where $n_{i\uparrow} + n_{i\downarrow} = 1$ 
independently of $i$, no matter what the rest of the parameters are. 
In comparison to the $\phi = 0$ diagrams which consist only 
of $\uparrow \downarrow$-PN, uniform-SF and polarised-SF regions, 
we find that the $\phi = 1/4$ diagrams have much richer structure involving 
large regions of stripe-ordered phases. To understand the physical origin of the
resultant phase diagrams and stripe order, next we discuss the analytically 
tractable high- and low-$h/g$ limits.

When $h/g$ is sufficiently high, we can directly read off the ground state 
of the $\sigma$ component from the HB for any given $\phi$. 
For $\phi = 1/4$, the energy spectrum consists of 4 bands: 
the $\sigma$ component is 
a $\sigma$-VAC for $\mu_\sigma \lesssim -2.83t$, 
a $\sigma$-N for $-2.83t \lesssim \mu_\sigma \lesssim -2.61t$,
a $\sigma$-I(1/4) for $-2.61t \lesssim \mu_\sigma \lesssim -1.082t$,
a $\sigma$-N for $-1.082t \lesssim \mu_\sigma \lesssim 1.082t$,
a $\sigma$-I(3/4) for $1.082t \lesssim \mu_\sigma \lesssim 2.61t$,
a $\sigma$-N for $2.61t \lesssim \mu_\sigma \lesssim 2.83t$ and
a $\sigma$-I(1/1) for $2.83t \lesssim \mu_\sigma$.
Using $\mu_\uparrow = \mu + h$ and $\mu_\downarrow = \mu - h$, 
the high-$h/g$ structure of Fig.~\ref{fig:pq14} immediately follows.
As $h/g$ gets smaller, the I and N phases must pave the way to 
ordered many-body ones, as increasing the strength of the pairing 
energy eventually makes them energetically less favourable. 
For $\phi = 0$, we confirmed that the $\uparrow \downarrow$-PN 
to polarised-SF phase transition boundary $g(h_c)$ is a monotonic function 
of $h$, which is simply because the non-interacting system has a very 
simple band structure with cosine dispersions. 
However, since the density of single-particle states is dramatically affected 
by the fractal band structure, the transition boundary 
$g(h_c)$ becomes a complicated function of $h$ for finite $\phi$. 
For instance, we find a sizeable hump in Fig.~\ref{fig:pq14}(a) 
around $h \approx 2.7t$ and another one in Fig.~\ref{fig:pq14}(b) around 
$h \approx 1.7t$, the peak locations of which coincide intuitively with the 
$\uparrow \downarrow$-PN regions that are sandwiched between 
VAC and/or I. 

On the other hand, when $h/g$ is sufficiently small, the ground state is 
expected to be an ordered many-body phase with no polarisation. 
In sharp contrast to the $\phi = 0$ case where uniform-SF is stable for any 
$\mu$, we show in Fig.~\ref{fig:pq14} that striped-SF and striped-SS are, 
respectively, stable for $\mu = 0$ and $\mu = -t$ when $\phi = 1/4$. 
Note that since $\mu = 0$ corresponds to half filling for any $\phi$, 
the unpolarised ground states necessarily have uniform fillings, 
\textit{i.e.}, $n_{i\uparrow} = n_{i\downarrow} = 1/2$. 
Therefore, in the low-$h/g$ limit, while only $|\Delta_i|$ is allowed to 
have spatial modulations in Fig.~\ref{fig:pq14}(a), both $|\Delta_i|$ 
and $n_{i \sigma}$ modulates in Fig.~\ref{fig:pq14}(b). 

The stripe order is a result of the HB: for a given $\phi$, 
the spectrum consists of $q$-bands in the 1st magnetic Brillouin zone 
within which each $\mathbf{k}$ state is $q$-fold degenerate. 
Therefore, not only intra- and inter-band 
pairings but also pairing with both zero and a set of non-zero center-of-mass 
momenta are allowed~\cite{zhai10, wei12}, leading to a 
non-uniform $|\Delta_i|$ with spatially-periodic modulations, 
\textit{e.g.}, a PDW order~\cite{agterberg08}. 
The directions of center-of-mass momenta determine the direction 
of modulations, making it gauge dependent, 
\textit{e.g.}, $y$ direction in Fig.~\ref{fig:S}.
When the striped-PDW order is sufficiently large, it drives an additional 
striped-CDW order in the total fermion filling, giving rise to 
striped-SS phases. We emphasise that the instability towards 
striped-PDW phases discussed in this paper is driven by $\phi \ne 0$ 
even when $h = 0$, and they cannot formally be identified with the 
FFLO-like non-striped PDW phases which are driven by $h \ne 0$ and 
are characterised by cosine-like sign-changing $|\Delta_i|$ oscillations 
along a spontaneously-chosen direction. 

It is clearly the cooperation between $\phi$ and $g$ that is responsible
for the broken spatial symmetry and appearance of stripe order, 
causing much more prominent stripes for intermediate $g$ at a given $h$. 
Depending on whether $q$ is odd or even, $|\Delta_i|$ modulation has 
a spatial period of $q$ or $q/2$ lattice sites, respectively.
The stripe order gradually fades away with increasing $g$, however, 
it survives even in the $g \gg W$ limit with $W$ the single-particle energy 
bandwidth, as long as $g/t$ is finite. 
Note in this limit that the physics is eventually determined 
by the two-body bound states, \textit{i.e.}, Cooper pairs become 
bosonic dimers, and unless $g/t\to \infty$, the dimer-dimer 
interaction ($g_{dd} \sim t^2/g$) is finite. Such weakly-repulsive dimers can 
effectively be described by the Hofstadter-Bose-Hubbard model, 
where superfluidity has recently been shown to break translation symmetry 
in the weakly-interacting limit~\cite{powell10}. 

In fact, all of our numerical results fit quite well with 
\begin{align}
\label{eqn:fit}
|\Delta_i| = |\Delta_0| + |\Delta_1| \cos(4\pi \phi i_y/\ell + \varphi),
\end{align}
in the entire unpolarised region (U-SF, S-SF and S-SS). 
Here, $|\Delta_0| \approx (g/2-4t^2/g)\sqrt{n(2-n)}$ for any $\mu$
determined by the total average filling $n$,
$|\Delta_1| \approx t^2/g$ for $\mu \approx 0$ (which becomes exact
only for $\mu = 0$ in the $g/t \to \infty$ limit), 
$i_y$ is the $y$ coordinate of site $i$, and
$\varphi$ is a constant phase shift set by the origin.
The microscopic origin of this expression can be best understood in 
the ideal-dimer limit, where $t_d \approx 2t^2/g$ and $\phi_d = p_d/q_d$, 
respectively, are their effective hopping and gauge field. 
Here, $p_d = 2p$ $(p)$ and $q_d = q$ $(q/2)$ for odd (even) $q$. 
Since the HB for dimers is $q_d$-fold degenerate, their ground state 
has contributions from all degenerate 
$k_{yd} = \{ 0, \pm 2\pi \phi_d f/\ell \}$ momenta where $f = 1, \cdots, q_d-1$, 
such that
$
\Psi_{id} = c_0 + \sum_{f} c_f \cos(2\pi \phi_d f i_y / \ell),
$
where $c_f$ are complex variational parameters. 
However, unlike atomic bosons where all of the degenerate states have 
equal weight, dimer bosons are fermion pairs and the number of ways of 
creating them with $k_{yd} = k_{y \uparrow} + k_{y \downarrow}$ 
momentum depends on $f$ and $\phi$, \textit{e.g.}, there 
are $2(q-f-1)+1$ ways of intra-band pairing when $q$ is even.
Thus, this analysis show that higher $|k_{yd}|$ states contribute less 
and less, forming a perturbative series. In addition, our fit Eq.~(\ref{eqn:fit}) 
suggests that the first order ($f = 1$) correction is already much smaller 
than the zeroth order ($f = 0$) one, and that the $f \ge 2$ terms are 
negligible. 

When $\phi$ is increased from $0$, we find that the transition from an
unpolarised to a polarised ordered phase occurs at a lower $h$ for any 
given $g$. This is a consequence of smaller $W$: as $\phi$ increases 
from $0$ to $1/6$ to $1/4$ then $W$ shrinks from $8t$ to $6.15t$ to $5.65t$, 
making it possible to polarise the ground state with a smaller and 
smaller $h$. In Fig.~\ref{fig:pq14}, the polarised region is dominated 
mainly by what we call the polarised-SF phase which is characterised by 
striped and/or non-striped PDW and/or SDW orders.

\begin{figure}[htb]
\centerline{\scalebox{0.6}{\includegraphics{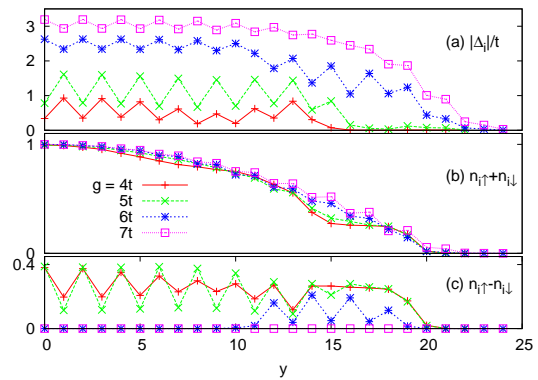}}}
\caption{\label{fig:trap14} (Color online)  
The trap profiles are shown for a cut along the $y$ direction (in units of $\ell$) 
when $x = 0$, $p/q = 1/4$, $\mu = 0$ and $h = t$.
}
\end{figure}

\textit{Trapped Systems.}
Next, we assume a harmonic confinement and choose $V_i = \alpha |\mathbf{r_i}|^2$ 
with strength $\alpha = 0.01t/\ell^2$.  For illustration purposes, in Fig.~\ref{fig:trap14} 
we show typical trap profiles for a cut along the $y$ direction when $x = 0$, 
$\phi = 1/4$, $\mu = 0$ and $h = t$.
Note that the system is completely unpolarised for $g = 7t$ with no SDW order, 
which is consistent with the phase diagrams shown in Fig.~\ref{fig:pq14}.
When $g/t$ is sufficiently small, the mini-gaps of the HB 
give rise to a wedding-cake structure with spatially-flat $n_{i \sigma}$ regions at 
integer multiples of $1/4$ fillings, but only the $n_{i\uparrow} \approx 0.25$ 
and $n_{i \downarrow} \approx 0$ region clearly survives in these figures.
In addition, while the CDW order is mostly washed out and barely visible
for $g = 6t$ and $g = 7t$, the PDW and SDW orders are large and conspicuous
in these figures (and can be much larger depending on the parameters
especially for large $q$), suggesting that PDW and/or SDW features may 
furnish clearest and direct evidence in trapped systems. 
It is also pleasing to see that the valleys of the PDW and CDW orders and 
peaks of the SDW order coincide when they coexist.

\textit{Conclusions.}
To summarise, motivated by the thrive of cold-atom experiments with 
artificial gauge fields, we analysed ground-state phases of the attractive 
Hofstadter-Hubbard model for both thermodynamic and trapped systems. 
Our main finding is that the interplay between the HB 
and SF order breaks spatial symmetry, and that the phase diagrams are 
dominated by stripe-ordered SF and SS phases which can be distinguished 
by their coexisting PDW, CDW and/or SDW orders. 
Such PDW superconductivity is relevant in a diverse range of 
systems~\cite{agterberg08, casalbuoni04, anglani14}, and given our promising results 
for atomic Fermi gases, we encourage further research in this direction 
with different lattice geometries, gauge fields, etc., in particular the beyond 
mean-field ones.

\begin{acknowledgments}
\textit{Acknowledgments.}
This work is supported by the Marie Curie IRG Grant No. FP7-PEOPLE-IRG-2010-268239, 
T\"{U}B$\dot{\mathrm{I}}$TAK 1001-114F232.
The author thanks A. L. Suba\c{s}{\i} and R. O. Umucal{\i}lar for discussions.
\end{acknowledgments}


\begin{thebibliography}{99}
\bibitem{zhaiweinote} Although the VL instability in the 2D Hofstadter-Hubbard lattice 
model is known for a long time (see, for instance~\cite{zhai10}), a cosine-shaped PDW 
instability has recently been found in a somewhat related model (a continuum 3D 
system in the lowest-Landau-level limit with an anisotropic trap)~\cite{wei12}. 
While our striped-PDW order is a direct result of the HB spectrum, the PDW 
instability that is found in Ref.~\cite{wei12} is driven by the specific form of the 
effective 1D interaction obtained after mapping the system to an effectively 
1D one in $\mathbf{k}$-space.
\bibitem{Hartree} We neglected them because not only the the self-consistent solutions 
converge much faster, the solutions are also much easier to interpret. In addition, 
since none of the PDW, CDW, SDW and VL instabilities are driven by these shifts, 
our mean-field results already paves the way to quantitative understanding of the 
possible ground states of the model.
\bibitem{ms}  However, there is no multiple solution problem in the low-$h/g$ limit, where 
we find unique U-SF, S-SF or S-SS solutions depending on whether $\mu = 0$ or not.
\end{thebibliography}
\end{document}